\documentclass[twocolumn,aps,prl,amsmath,amssymb]{revtex4}

\usepackage{graphicx}

\usepackage{dcolumn}

\usepackage{bm}

\DeclareGraphicsExtensions{.pdf,.eps,.png,.jpg,.mps}

\begin{document}

\preprint{AIP/123-QED}

\title{Towards Substrate Engineering of Graphene-Silicon Schottky Diode Photodetectors}

\author{H. Selvi$^{1}$}

\author{N. Unsuree$^{1}$}

\author{E. Whittaker$^{1,3}$}

\author{M.P. Halsall$^{1,3}$}

\author{E.W. Hill$^{2,4}$}

\author{P. Parkinson$^{3,5}$}

\author{T.J. Echtermeyer$^{1,2,3}$}

\email{tim.echtermeyer@manchester.ac.uk}

\affiliation{$^1$School of Electrical \& Electronic Engineering, University of Manchester, Manchester M13 9PL, UK}

\affiliation{$^2$National Graphene Institute, University of Manchester, Manchester M13 9PL, UK}

\affiliation{$^3$Photon Science Institute, University of Manchester, Manchester M13 9PL, UK}

\affiliation{$^4$Manchester Centre For Mesoscience and Nanotechnology, University of Manchester, Manchester M13 9PL, UK}

\affiliation{$^5$School of Physics and Astronomy, University of Manchester, Manchester M13 9PL, UK}

\begin{abstract}

Graphene-Silicon Schottky diode photodetectors possess beneficial properties such as high responsivities and detectivities, broad spectral wavelength operation and high operating speeds. Various routes and architectures have been employed in the past to fabricate devices. Devices are commonly based on the removal of the silicon-oxide layer on the surface of silicon by wet-etching before deposition of graphene on top of silicon to form the graphene-silicon Schottky junction. In this work, we systematically investigate the influence of the interfacial oxide layer, the fabrication technique employed and the silicon substrate on the light detection capabilities of graphene-silicon Schottky diode photodetectors. The properties of devices are investigated over a broad wavelength range from near-UV to short-/mid-infrared radiation, radiation intensities covering over five orders of magnitude as well as the suitability of devices for high speed operation. Results show that the interfacial layer, depending on the required application, is in fact beneficial to enhance the photodetection properties of such devices. Further, we demonstrate the influence of the silicon substrate on the spectral response and operating speed. Fabricated devices operate over a broad spectral wavelength range from the near-UV to the short-/mid-infrared (thermal) wavelength regime, exhibit high photovoltage responses approaching 10$^6$ V/W and short rise- and fall-times of tens of nanoseconds.

\end{abstract}

\maketitle
Graphene is an appealing material for ultrafast and broadband photodetection applications due to its high charge carrier mobility \cite{zhang2005experimental,geim2009graphene} and ultra-wide spectral absorption range \cite{mak2012optical,li2008dirac}. The initial examples of graphene photodetectors are mostly based on metal-graphene (MG) junctions \cite{mueller2010graphene,xia2009ultrafast,lee2008contact} and graphene p-n junction architectures \cite{xu2009photo,gabor2011hot,lemme2011gate}. Despite their broadband operation at ultrafast speeds \cite{pospischil2013cmos}, they generally exhibit low responsivities (limited to a few mAW$^{-1}$) due to the intrinsically low optical absorption of monolayer graphene (2.3\%) \cite{nair2008fine}. Further, the small photoactive area limits their use for real-world applications \cite{gabor2011hot,lemme2011gate,mueller2009role}. Recently, there is a surge of interest in using graphene to replace the metal electrode on semiconductor surfaces to realize Schottky diodes \cite{di2016graphene,li2016graphene,tongay2009graphite,dragoman2010graphene,chen2011graphene,tongay2012rectification,chen2012gate,yang2012graphene,yim2013characterization,sinha2014ideal,parui2014temperature,amirmazlaghani2013graphene,wang2013high,an2013tunable,lv2013high,an2013metal,liu2014quantum,chen2015high,goykhman2016chip,li2016high,srisonphan2016hybrid,di2016tunable,riazimehr2016spectral,wan2017self,shen2017high,riazimehr2017graphene,di2017hybrid,tao2017hybrid,li2015carbon,li2010graphene,miao2012high,an2013optimizing,lin2013graphene,song2015role,kim2013chemically,fattah2014graphene,singh2014tunable,uddin2014functionalized,zhu2015photo}. Unlike conventional bulk metals, graphene's high optical transmittance \cite{nair2008fine}, tuneable Fermi level \cite{yu2009tuning}, high mobility of charge carriers \cite{zhang2005experimental} and atomically thin structure \cite{novoselov2005two} potentially bring additional functionalities to Schottky diode platforms. They can be used in a variety of applications such as photodetection \cite{amirmazlaghani2013graphene,wang2013high,an2013tunable,lv2013high,an2013metal,liu2014quantum,chen2015high,goykhman2016chip,li2016high,srisonphan2016hybrid,di2016tunable,riazimehr2016spectral,wan2017self,shen2017high,riazimehr2017graphene,di2017hybrid,tao2017hybrid} solar energy harvesting \cite{li2015carbon,li2010graphene,miao2012high,an2013optimizing,lin2013graphene,song2015role}, chemical gas sensing \cite{kim2013chemically,fattah2014graphene,singh2014tunable,uddin2014functionalized,zhu2015photo} and current rectification \cite{dragoman2010graphene}. A further benefit of the graphene-silicon Schottky platform is its simple architecture, suitability for large scale fabrication and the potential for integration into the back end-of-line (BEOL) CMOS processing \cite{pospischil2013cmos,pasternak2016graphene}.
Particularly in photodetection applications, the G-Si diode provides an efficient hybrid platform where both graphene and silicon can be used as absorbing materials for different wavelength ranges \cite{riazimehr2016spectral}. Devices shows high responsivities comparable to commercial silicon photodiodes for wavelength ranges with photon energies above the silicon bandgap which is enabled by the high optical transmittance of graphene of more than 97\% \cite{an2013tunable,riazimehr2016spectral}. Detection of radiation with energies below the silicon band gap is enabled by the broadband absorption of graphene \cite{amirmazlaghani2013graphene,wang2013high}. Responsivities reduce to a few mAW$^{-1}$ \cite{amirmazlaghani2013graphene} and less for longer wavelengths \cite{riazimehr2016spectral} which is comparable to the responsivities of planar MG junction PDs \cite{echtermeyer2014photothermoelectric}. However, compared to MG junction PDs the G-Si Schottky diode offers the advantage of large photoactive area, formed by the whole lateral junction surface.   
Optical and electrical characteristics of planar G-Si Schottky diodes have been comprehensively investigated in earlier reports. Differences in the device characteristics such as ideality factor, level of dark current and spectral dependent photo-response are mostly attributed to profound effects of interface properties, choice of materials and fabrication route in the past literature \cite{parui2014temperature,riazimehr2016spectral,wan2017self,shen2017high,di2017hybrid}. In this regard, the effect of interfacial oxide layer on rectification characteristics \cite{li2016high}, and solar cell efficiency \cite{song2015role}, as well as influence of graphene doping on spectral response \cite{an2013tunable} and on the efficiency of solar cells \cite{miao2012high} were investigated in more specific reports. However, the influence of substrate properties on optoelectronic characteristics of the diode has not been discussed in detail before. 
In this work, we characterised G-Si diodes conducting current-voltage (I-V) and high speed optical measurements over broad wavelength and light intensity ranges to investigate the effects of interface and substrate properties on the photoresponse and response time characteristics of graphene-silicon Schottky diodes.

\begin{figure}[htbp]
\centering{
\includegraphics[width=80mm]{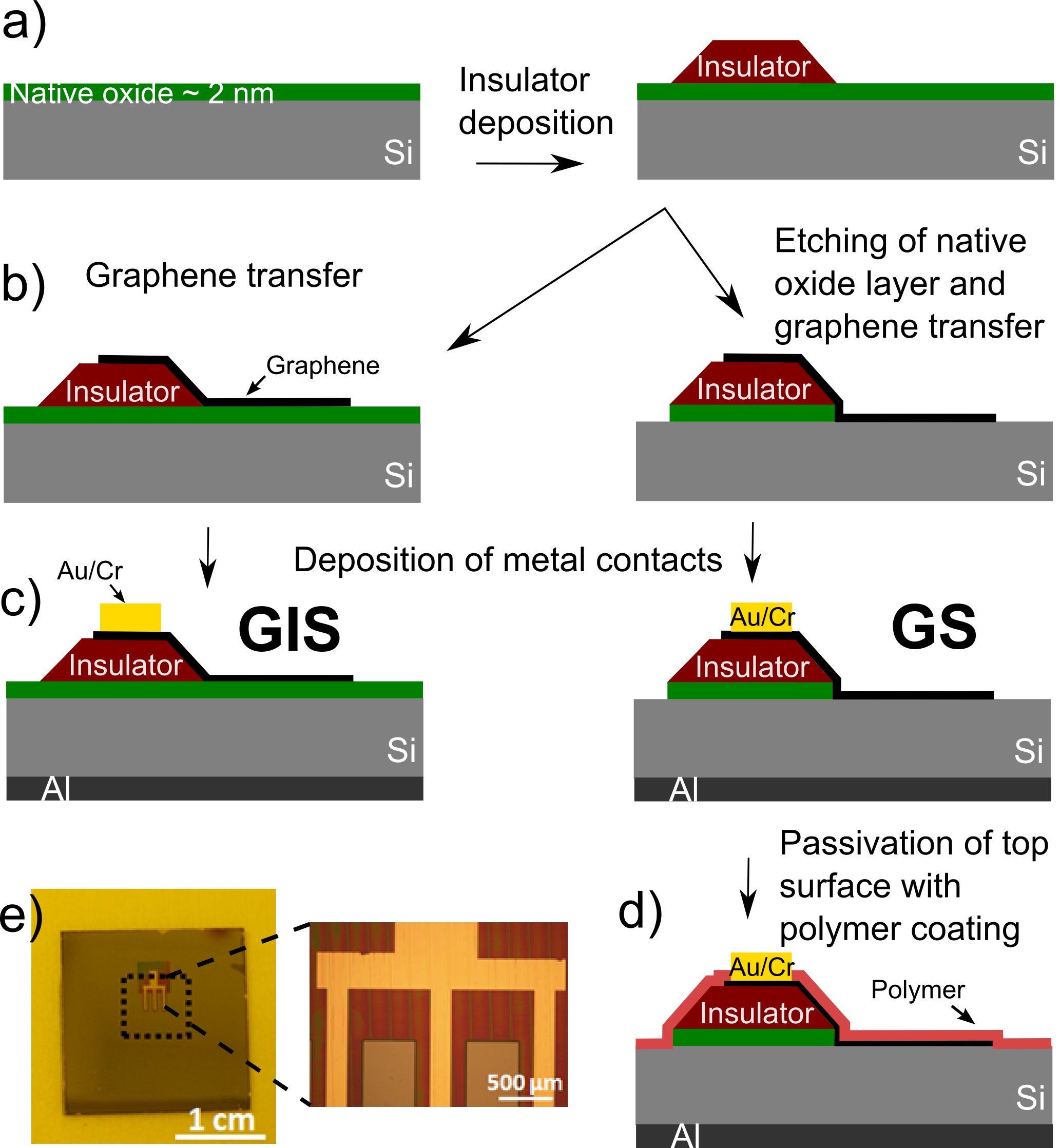}}
\caption{Fabrication of graphene/silicon Schottky diodes. a) Plain silicon with native oxide layer is used as starting substrate and and insulating spacer is defined. b) A split is performed to remove the native oxide layer for the GS devices and graphene is transferred onto the substrate. c) Subsequently, the contacts are deposited. d) Passivation of the GS device to delay re-growth of the interfacial layer. e) Image of a fabricated device.}
\label{fig:process}
\end{figure}

Two different sets of devices were fabricated, Graphene/Silicon (GS) and Graphene/\allowbreak Insulator/Silicon (GIS) Schottky diodes. The process flow is depicted in fig.\ref{fig:process}. Bare n-doped silicon wafers (specification $\rho$ = 1-10 Ohm cm, extracted doping level N$_{\rm{d}}$ $\approx$ 3.5$\times10^{14}$ cm$^{-3}$) with native silicon oxide layer (SiO$_2$) have been used as substrate. Conventionally, wafers with grown or deposited silicon oxide of greater thickness in the order of tens to hundreds of nanometers and subsequent etching of this oxide layer are used to fabricate devices. Our employed fabrication route allows especially in the case of the GIS devices an initially homogeneous natural oxide layer of well defined thickness of $\approx$ 2nm \cite{morita1990growth}. An insulating spacer, serving later as base for the contact metallization to graphene, is fabricated from negative photoresist, defined by optical lithography and subsequently hard-baked to withstand further processing. The spacer has a thickness of 800nm to prevent dielectric breakdown for the range of bias voltages applied. Further, the exposure dose for optical lithography has been optimized to yield rounded edges of the spacer which facilitates transfer of the graphene layer on top of this structure in a later step and avoids tearing of the graphene layer. A split is then performed to fabricate GS and GIS devices. For the GS devices, a hydrofluoric acid (HF) dip is carried out to remove the native oxide layer while GIS devices are left untreated. Graphene grown by chemical vapour deposition (CVD) on copper foil is then transferred to both device types employing a process based on poly-methyl-methacrylate (PMMA). This transfer step is carried out immediately after the HF-dip for the GS device to minimize the re-growth of an interfacial oxide layer. A further lithography step is then carried out to define contacts to the graphene layer using gold/chromium (Au/Cr). Aluminium (Al) serves as low work function metallization to the silicon back side to form an ohmic contact. An additional layer of photoresist is spun onto the sample surface of the GS device to delay re-growth of an interfacial oxide layer by reducing exposure of the graphene-silicon junction to the ambient atmosphere. Fig.\ref{fig:process}e shows an image of a fabricated device. It is noteworthy that our simple process for the GIS device allows rapid fabrication of devices with high throughput. Particularly the absence of HF-etching simplifies the processing and further eliminates the need for depositing graphene on the silicon surface shortly after HF-etching. Devices have intentionally been designed to have a large junction area in the order of $\approx$ 60mm$^2$. The large lateral dimensions of the junction in the order of mm lead to a high vertical depletion region length of the junction in silicon ($\mu$m) ratio. This high ratio reduces fringe and edge effects of the electric field and promotes a 1-dimensional (vertical) electric field in the junction region which allows treating devices as simple parallel plate capacitor. The contact to graphene has been designed as finger-like structure (inset) with the aim to increase the perimeter of the contact and reduce contact resistance \cite{Smithcontact}.

\begin{figure}[htbp]
\centering{
\includegraphics[width=80mm]{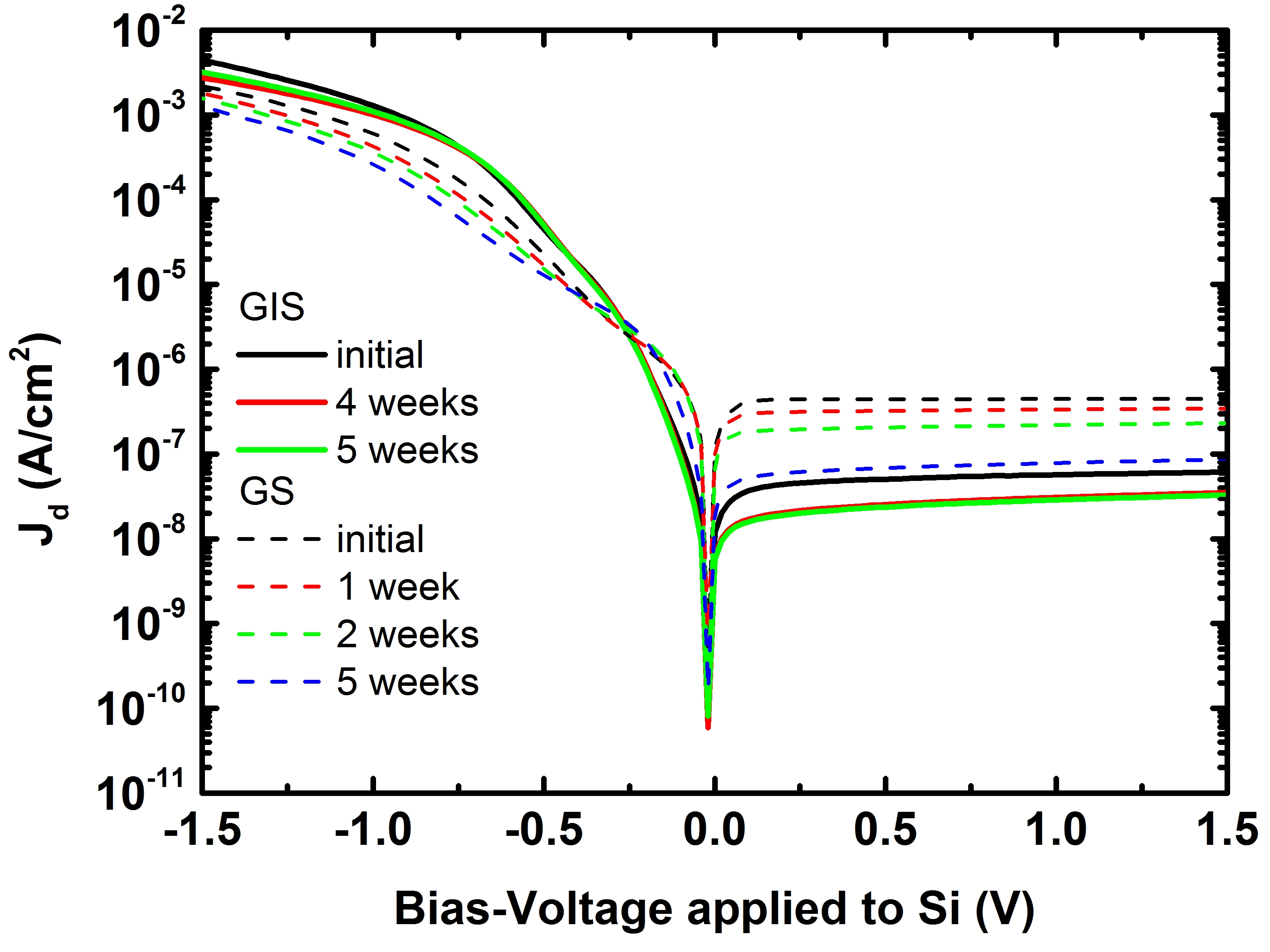}}
\caption{Current-voltage characteristics for the GS and GIS device at varying times after fabrication.}
\label{fig:IV}
\end{figure}

The dark current-voltage (IV) characteristics of both device types have been recorded immediately after device fabrication and repeatedly re-measured over the following five weeks. Fig.\ref{fig:IV} shows that the GIS device exhibits and on- to off-current ratio (I$_{\rm{on}}$/I$_{\rm{off}}$) of 10$^5$, defined by the ratio of currents at voltages of + and - 1.5V, respectively. In comparison, the GS device initially exhibits an I$_{\rm{on}}$/I$_{\rm{off}}$ ratio of 10$^4$. The Schottky-barrier-heights (SBH) determined by temperature dependent measurements are 0.62eV and 0.45eV for the GIS and GS device, respectively (see supplementary information). The built-in potential derived from CV measurements is $\phi_{\rm{bi}}$ = 0.58V for the GIS and $\phi_{\rm{bi}}$ = 0.31V for the GS device (see supplementary information). Both the GIS and GS device have a series resistance R$_{\rm{S}}$ of $\approx$ 470 $\Omega$ and ideality factors $\eta$ of 2.8 and 4.8, respectively (see supplementary information). In the forward biased regime for voltages in the range of $\approx -0.25..0V$, the GS device exhibits a much sharper increase in current with bias voltage than the GIS device. Both the higher I$_{\rm{on}}$/I$_{\rm{off}}$ ratio and the less sharp on-set in current of the GIS device can be attributed to the presence of the interfacial layer \cite{sze2007physics}. The natural oxide layer acts as a tunneling barrier and suppresses current flow for low voltage drops across the oxide layer \cite{sze2007physics}.

Devices have been stored in ambient atmosphere over several weeks and re-measured subsequently. With increasing lifetime of the devices, both forward and reverse current decrease in both devices. After five weeks, both forward and reverse currents are approximately reduced two-fold in the GIS device. This aging effect is much more pronounced in the GS device. The reverse current decreases by almost an order of magnitude and the I$_{\rm{on}}$/I$_{\rm{off}}$ ratio increases. Correspondingly, the SBH increases to 0.61 eV and the built-in potential to $\phi_{\rm{bi}}$ = 0.44V after five weeks. Further, it is visible that the GS device starts to follow the characteristics of the GIS device in the low voltage forward biased region, the slope of the forward current increase is reduced. These changes in the GS device can be attributed to the re-growth of the interfacial layer with increasing lifetime \cite{morita1990growth,li2016high}. The diffusion of oxygen and water through cracks and grain boundaries in CVD graphene leads to an oxidation of the silicon surface underneath \cite{li2016high,morita1990native}. A similar effect, even though less pronounced, occurs for the GIS device. This is counter-intuitive as the initially present oxide layer thickness is self-limiting, has been grown over months of wafer storage and should have reached its maximum thickness. The exact reason for this re-growth is unknown but it is highly likely that molecular species such as oxygen and water are trapped underneath the graphene layer which promotes the growth of SiO$_2$ \cite{Vasupressure}.

\begin{figure}[htbp]
\centering{
\includegraphics[width=80mm]{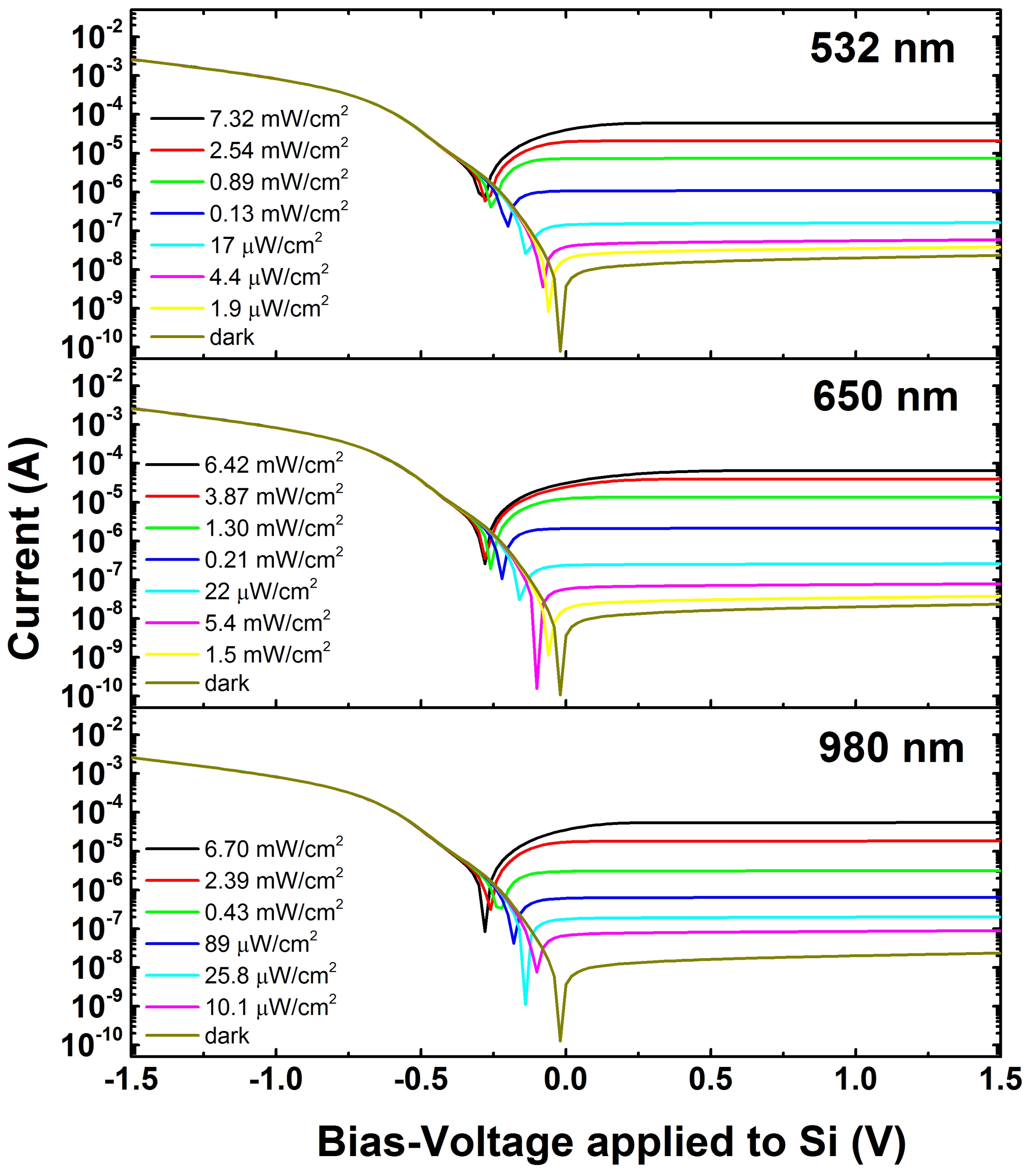}}
\caption{IV characteristics of the GIS device under illumination at different wavelengths and light intensities. a) 532nm, b) 650nm, c) 980nm.}
\label{fig:photo}
\end{figure}

Subsequently, we characterized both GIS and GS devices under illumination at different wavelengths and light intensities. Graphene-Silicon Schottky diode photodetectors exhibit two different operating regimes. In the wavelength region below $\lambda \approx$ 1.1$\mu$m, silicon is the main light absorber due to optical excitation of charge carriers above its band gap of $E_g = 1.1eV$. For longer wavelengths $\lambda$ \textgreater 1.1$\mu$m, optical absorption takes place in graphene and silicon is optically transparent at employed low doping level N$_d$ $\approx 3.5\times10^{14}$ cm$^{-3}$. Optically excited charge carriers in graphene gain sufficient energy to overcome the Schottky barrier formed at the graphene-silicon interface and lead to a photoresponse \cite{amirmazlaghani2013graphene,wang2013high}. For light detection in the visible (VIS) and near-infrared (NIR) wavelength region ($\lambda <$ 1.1$\mu$m) a large SBH is beneficial. It leads to reduced reverse current density, as demonstrated for the GIS diode (Fig.\ref{fig:IV}), and improves the signal-to-noise ratio (SNR). In contrary, for the generation of a photocurrent due to longer wavelength light a reduced SBH as in the GS device is required as it facilitates the transmission of optically excited charge carrier in graphene over the Schottky barrier. 

For opto-electronic characterization in the VIS and NIR wavelength regime, continuous wave (CW) laser sources with wavelengths of 532, 650 and 980nm at varying intensities as well as illumination from a broad band white light source filtered with a monochromator have been used. The photoresponse of the GIS device was determined by recording the IV characteristics under illumination and is shown in fig.\ref{fig:photo} for wavelengths of 532, 650 and 980nm. The reverse current increases with increasing light intensities under illumination for all wavelengths. With increasing light intensity, the open circuit voltage $V_{\rm{oc}}$, the voltage when the diode forward current and photocurrent compensate each other ($I_{\rm{forward}} + I_{\rm{photo}} = 0$, minimum in the IV curves), shifts towards the forward biased region of the diode. It is notable that independent of light intensity and wavelength, the current in the reversed biased region saturates sharply and does not increase with increased reverse bias. Further, higher light intensities lead to a stretch out of the IV curves; the voltage that needs to be applied beyond $V_{\rm{oc}}$ in the reverse biased regime to drive the device into saturation increases with higher light powers.

\begin{figure}[htbp]
\centering{
\includegraphics[width=80mm]{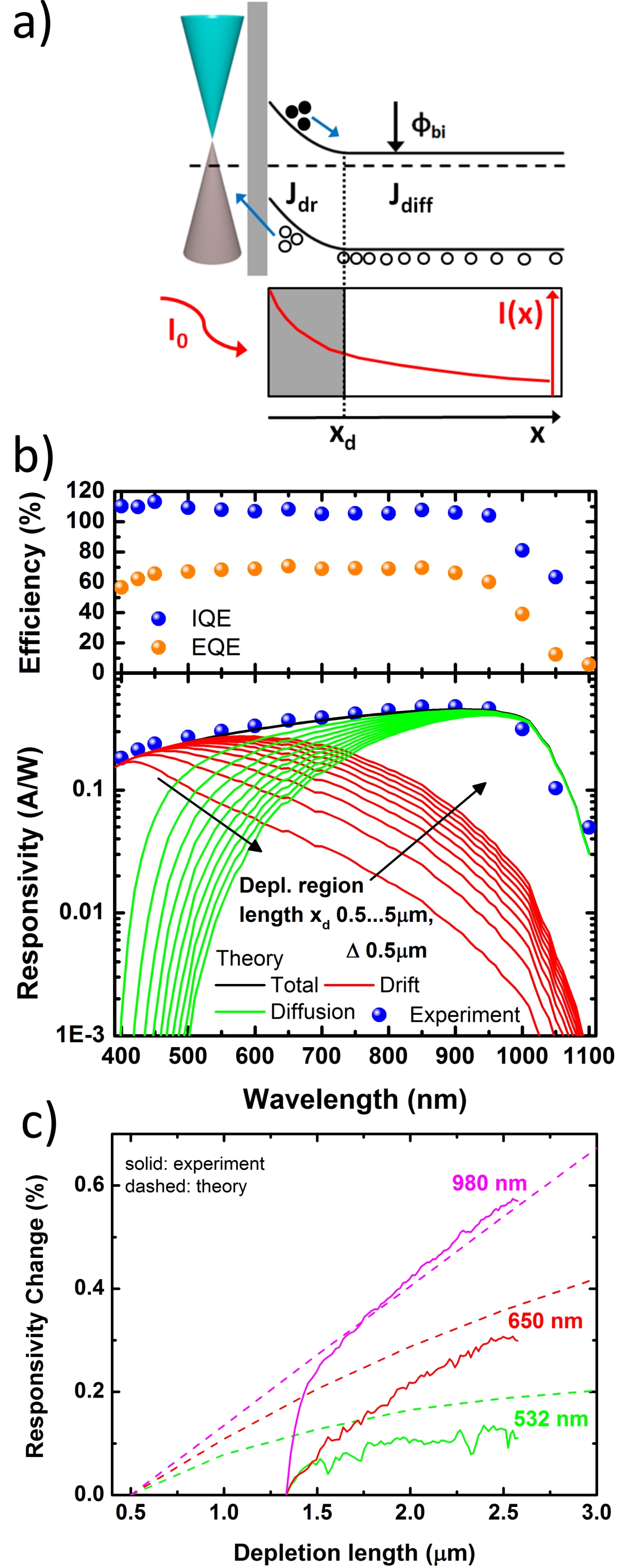}}
\caption{Comparison of theoretical and experimentally observed photoresponse in graphene-silicon Schottky diode photodetectors. a) Energy band diagram of the device and cross-section through the silicon substrate, indicating light and optical carrier generation as a function of depth and contributions of drift and diffusion current, J$_{\rm{dr}}$ and J$_{\rm{diff}}$, respectively. b) Theoretical spectral responsivity as a function of depletion length and contributions of drift and diffusion currents to the overall response in comparison with experimentally determined responsivities and efficiencies of the device. c) Percentage change in responsivity with increasing depletion length.}
\label{fig:theo}
\end{figure}

To understand observed opto-electronic properties of graphene-silicon Schottky diode photodetectors, it is instructive to review the processes involved in the photocurrent generation in such devices. Fig.\ref{fig:theo}a shows the band diagram and cross-section through the silicon substrate as a function of depth. When graphene is in contact with silicon, band bending occurs in silicon and a depletion region of length $x_{\rm{d}}$ forms \cite{sze2007physics,tyagi1984physics}. The built-in potential $\phi_{\rm{bi}}$ drops across the depletion region \cite{neamen2003semiconductor}. The depletion length in a Schottky diode depends on the applied reverse bias and can be calculated using the full depletion approximation as \cite{sze2007physics}

\begin{equation}
x_{\rm{d}} = \sqrt{\frac{2 \epsilon_0 \epsilon_{\rm{Si}}}{q N_{\rm{d}}} (\phi_{bi} + V_{\rm{b}})}
\label{eq:xd}
\end{equation}

with $\epsilon_0$ free space permittivity, $\epsilon_{\rm{Si}}$ relative permittivity of silicon, $q$ electron charge and $V_{\rm{b}}$ applied reverse bias.

Light absorption in silicon follows Beer-Lambert law, describing the exponential decay of light intensity $I$ as a function of depth $x$ \cite{hecht}. Taking into account optical reflection at the device surface it can be described by \cite{hecht,sze2007physics}

\begin{equation}
I(x) = I_0 \ (1-R) \ e^{-\alpha x}
\label{eq:lblaw}
\end{equation}

Here, $I_0$ is the incident light intensity, $\alpha$ the wavelength dependent absorption coefficient, and $R$ the wavelength dependent optical reflection coefficient at the device surface. This leads to a depth $x$ dependent generation rate of electron-hole pairs $G_{\rm{e,h}}$ in the silicon substrate \cite{sze2007physics}

\begin{equation}
G_{\rm{e,h}}(x) = \alpha \ \frac{\lambda}{hc} \ I_0 \ (1-R) \ e^{-\alpha x}
\label{eq:geh}
\end{equation}

with Planck constant $h$ and $c$ speed of light. A quantum efficiency of one is assumed in eq.\ref{eq:geh}, \textit{i.e.} every photon generates an electron-hole pair.

Optically excited charge carriers generate a photoresponse in the device due to two processes. Band bending and corresponding electric fields within the depletion region separate electron-hole pairs and lead to a drift current density $J_{\rm{dr}}$ (fig.\ref{fig:theo}a). Photo-generated holes (minority carriers) in the non-depleted bulk of the silicon substrate exhibit a concentration gradient as a function of depth (fig.\ref{fig:theo}a). This concentration gradient leads to a diffusion current density $J_{\rm{diff}}$. The total photo current of a graphene-silicon Schottky diode can be described in a similar way to a conventional pn-junction photodiode, yet, with omitted p-doped region \cite{sze2007physics}. The total current density, $J_{\rm{tot}}$ is given  by \cite{sze2007physics}

\begin{equation}
J_{\rm{tot}} = J_{\rm{dr}} + J_{\rm{diff}}
\label{eq:jtot}
\end{equation}

with drift current density

\begin{equation}
J_{\rm{dr}} = q \ I_{0} \ (1-R) \ \frac{\lambda}{hc} \ (1-e^{-\alpha x_{\rm{d}}})
\label{eq:jdr}
\end{equation}

and diffusion current density

\begin{multline} 
J_{\rm{diff}} = q \ I_{0} \ (1-R) \ \frac{\lambda}{hc} \ \frac{\alpha L_{\rm{p}}}{\alpha^2 L_p^2-1} \ e^{-\alpha x_{\rm{d}}}\\ 
\left\{ \alpha L_{\rm{p}} - \frac{S_{\rm{p}} L_{\rm{p}}/D_{\rm{p}} \ \left[\cosh(H'/L_{\rm{p}})-e^{-\alpha H'}\right]}{S_{\rm{p}} L_{\rm{p}}/D_{\rm{p}} \sinh(H'/L_{\rm{p}}) + \cosh(H'/L_{\rm{p}})} \right.\\
\left. -\frac{\sinh(H'/L_{\rm{p}}) + \alpha L_{\rm{p}} e^{-\alpha H'}}{S_p L_{\rm{p}}/D_{\rm{p}} \sinh(H'/L_{\rm{p}}) + \cosh(H'/L_{\rm{p}})} \right\}
\label{eq:jdiff}
\end{multline}

Here, $q$ is the electron charge, $h$ Planck's constant, $c$ the speed of light, $L_{\rm{p}}$ the diffusion length of holes in the n-doped substrate, $S_{\rm{p}}$ the recombination velocity of carriers at the back-side of the substrate, and $D_{\rm{p}}$ the diffusion coefficient. $H' = H - x_{\rm{d}}$, with $H$ the physical thickness of the silicon substrate, denotes the length of the intrinsic substrate region. We explicitly consider the substrate thickness in our calculation of the diffusion current. Hole diffusion lengths $L_{\rm{p}}$ in low doped silicon are in the order of hundreds of $\mu$m \cite{sze2007physics} and comparable to the silicon substrate thickness of typically $H$ = 500 $\mu$m. Especially for longer wavelengths and corresponding low absorption coefficient $\alpha$, light is able to penetrate deep into the silicon substrate and leads to non-negligible diffusion currents.

Fig.\ref{fig:theo}b shows the calculated wavelength dependent responsivity $\mathcal{R}$  based on equations \ref{eq:jdr} and \ref{eq:jdiff} as a function of depletion region length. The diffusion coefficient has been derived from the Einstein relation as $D_{\rm{p}} = \mu_{\rm{p}} \frac{kT}{q}$ based on a hole mobility of $\mu_{\rm{p}} = 400 \ \frac{cm^2}{Vs}$ for low doped silicon substrates at room temperature \cite{sze2007physics}. The diffusion length has been derived from the recombination time $\tau_{\rm{p}} = 2\times10^{-4}s$ for n-doped silicon with a doping level of N$_d$ = 3.5$\times10^{14}$ cm$^{-3}$ \cite{sze2007physics} and is calculated as $L_{\rm{p}} = \sqrt{D_{\rm{p}} \tau_{\rm{p}}} = 450 \ \mu m$. $S_{\rm{p}} = \infty$ has been used and reflection $R$ and absorption coefficient $\alpha$ have been calculated from the complex refractive indices of silicon \cite{palik}. No fitting of any parameters has been used. Both drift and diffusion currents contribute equally in magnitude to the total photoresponse. Light of shorter wavelengths is absorbed closer to the silicon surface and the drift current contribution dominates in this wavelength range. Longer wavelength light penetrates deeper into the substrate and leads to diffusion currents exceeding the drift currents.

With increasing depletion region length, the spectral response of the drift current contribution is shifted towards longer wavelengths and its absolute contribution increases. Similarly, the diffusion current contribution shifts towards longer wavelengths with increasing depletion region length. The magnitude of the total photoresponse of the device is largely independent of depletion region length since drift and diffusion current contributions compensate each other. However, the ratio of drift to diffusion currents is strongly dependent on the depletion region length. Especially for high operating speeds, it is desirable to suppress diffusion currents due to long recombination times of charge carriers involved in this process which limits response times. For comparison, the experimentally determined photocurrent responsivity $\mathcal{R}_c$ at $V_b = 1.5V$ has been included, calculated as $\mathcal{R}_{\rm{c}} = I_{\rm{Photo}} / P_{\rm{Light}}$ with $I_{\rm{Photo}} = I_{\rm{Illuminated}} - I_{\rm{dark}}$ and $P_{\rm{Light}}$ the incident, external light power. The experimental and theoretical spectral dependence of the responsivities are in good agreement. The internal quantum efficiency (IQE), determined as the ratio of experimentally determined responsivity to theoretical responsivity, based on the ideal number of photons absorbed, reaches values greater than 100\% for wavelengths up to 950nm. It should be noted that the IQE for $\lambda$ = 1100nm has been omitted due to uncertainty in the absorption coefficient $\alpha$ close to the absorption edge of silicon and would lead to IQE values greater than 100\%. The external quantum efficiency $EQE$ is calculated from the experimental photocurrent responsivity $\mathcal{R}_{\rm{c}}$ as

\begin{equation}
EQE = \frac{\mathcal{R}_{\rm{c}}}{q} \ \frac{hc}{\lambda}
\label{eq:eqe}
\end{equation}

The EQE reaches values of 60-70\% in the wavelength range from 400-950nm after which it decays for longer wavelengths, approaching photon energies close to the band gap of silicon where silicon is not light absorbing anymore. Remarkably, the EQE is almost constant in the wavelength range from 400-950nm. It does not decrease for shorter wavelengths, since the formed Schottky junction in this device is located just beneath the silicon surface as opposed to conventional pn-junction photodiodes based on a buried junction \cite{sze2007physics}. The location of the Schottky junction just below the surface promotes the conversion of shorter wavelength light which is strongly absorbed just below the surface into a photocurrent. The high optical transparency of graphene enhances the efficient transduction of short wavelength light into a photoresponse compared to metals employed as electrode in Schottky photodiodes. For example, gold becomes strongly light absorbing for wavelengths shorter than $\approx$ 500nm \cite{palik}.

Further, the observed strong saturation of the photocurrent with increasing reverse bias (fig.\ref{fig:photo}) is reflected by equations \ref{eq:jdr} and \ref{eq:jdiff}. Fig.\ref{fig:theo}c shows the theoretical and experimental percentage change in photocurrent with increasing depletion region length for wavelength of $\lambda$ = 532, 650 and 980nm. The small overall increase of less than one percent of the photocurrent, and the wavelength dependence are in agreement for theory and experiment. Therefore we argue that observed saturation of the photocurrent, \textit{i.e.} its independence of reverse bias voltage, is caused by the silicon substrate and not limited by the available states within graphene that can be occupied by photo generated charge carriers as claimed in reference \cite{an2013tunable}.

\begin{figure*}[htbp]
\centering{
\includegraphics[width=175mm]{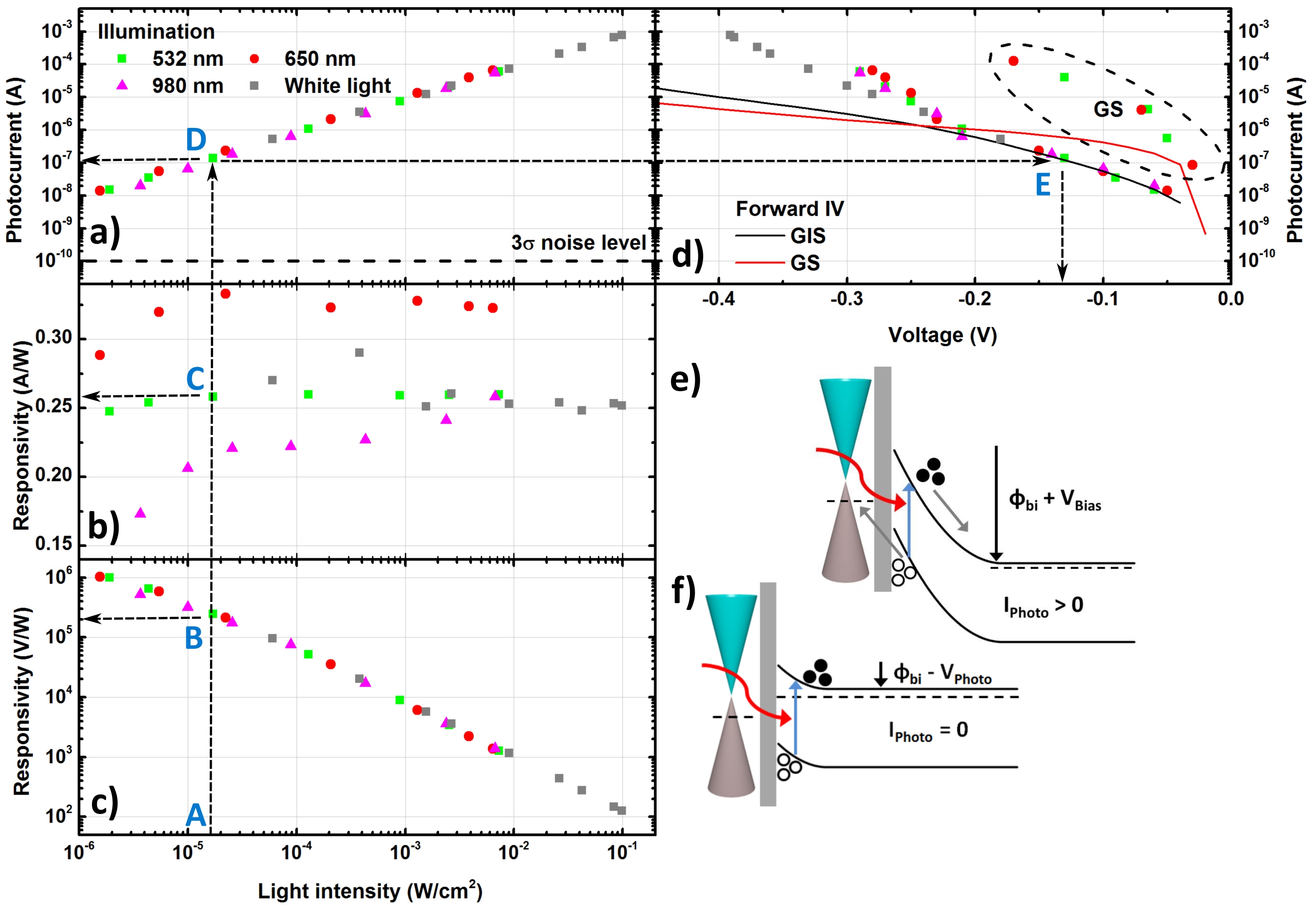}}
\caption{Comprehensive overview of the fundamental parameters of the GIS device with dependence of wavelength and light power. a) Absolute photocurrent. b) Photocurrent responsivity at a reverse bias of $V_{\rm{b}}$ = 1.5V. c) Open circuit (I=0) photovoltage responsivity. d) Absolute open circuit voltage $V_{\rm{oc}}$ with comparison to the GS device and forward currents of the diodes. e,f) Schematic energy band diagram of a Schottky diode photodetector at applied reverse bias (photocurrent mode) and under open circuit conditions (photovoltage mode), respectively.}
\label{fig:pdpar}
\end{figure*}

A comprehensive overview of the fundamental properties of the GIS device for various light wavelengths, obtained with laser and white light sources at different optical intensities is shown in fig.\ref{fig:pdpar}. The four linked panels of a) absolute photocurrent, b) photocurrent responsivity $\mathcal{R}_{\rm{c}}$, c) photovoltage responsivity $\mathcal{R}_{\rm{v}}$ and d) open circuit voltage $V_{\rm{oc}}$ give a coherent overview of the fundamental properties of the device and allow determination of the optimum operating conditions for a desired application. This is indicated by the arrows and blue letters. For example, incident light with an intensity in the order of $I_0$ = 20$\mu$W/cm$^2$ (fig.\ref{fig:pdpar}c, point A) will lead to an absolute photocurrent in the order of $I_{\rm{pc}}$ = 100nA (panel a, point D). Panel a) further shows the linear trend of the generated absolute photocurrent as a function of light intensity. The dark current noise floor of the GIS device has been included and its standard deviation of 3$\sigma$ = $10^{-10}$A, two orders of magnitude below the photocurrents recorded in our experiments, indicates further potential of GIS devices for low light intensity detection. The according photocurrent responsivity $\mathcal{R}_{\rm{c}}$ in A/W at a reverse bias voltage of $V_{\rm{b}}$ = 1.5V can be derived from fig.\ref{fig:pdpar}b, point C with the corresponding band diagram shown in panel e). As described in Fig.\ref{fig:theo}b) and according to eq.\ref{eq:jdr} and \ref{eq:jdiff}, the photocurrent responsivity $\mathcal{R}_{\rm{c}}$ is wavelength dependent. Further, photocurrent responsivity $\mathcal{R}_{\rm{c}}$ decreases with decreasing light powers which is not straight forward visible in the log-log plot in panel a). 

The photovoltage responsivity $\mathcal{R}_{\rm{v}} = V_{\rm{oc}} / P_{\rm{Light}}$ in V/W for a given light intensity can be determined from fig.\ref{fig:pdpar}c with the band diagram shown in fig.\ref{fig:pdpar}f. While photocurrent responsivity $\mathcal{R}_{\rm{c}}$ decreases for lower light intensities and remains of the same order of magnitude in the range of light intensities used in the experiments, photovoltage responsivity $\mathcal{R}_{\rm{v}}$ exhibits a more pronounced dependence on light intensity. It increases by four orders of magnitude approaching $10^6$ V/W when light intensities are decreased to the $\mu$W/cm$^2$ range.

Fig.\ref{fig:pdpar}d shows the absolute open-circuit photovoltage for the GIS device. For comparison, the open-circuit photovoltage for the GS device has been included. Forward dark IV curves are plotted additionally for both devices as the open-circuit voltage is determined by when dark forward current and photocurrent compensate each other. In the GIS device, the open-circuit voltage follows the dark IV curve for low light intensities up to $\approx$ 0.22V before it deviates and approaches saturation, determined by the built-in voltage $\phi_{\rm{bi}}$. On the contrary, the GS device exhibits a reduced open-circuit voltage compared to the GIS device for an identical photocurrent.

The reduced photocurrent responsivity $\mathcal{R}_{\rm{c}}$ (fig.\ref{fig:pdpar}b) and increased photovoltage responsivity $\mathcal{R}_{\rm{v}}$ (fig.\ref{fig:pdpar}c) for low light intensities can be qualitatively explained due to the presence of the interfacial oxide layer in the GIS device according to the suggestion in \cite{song2015role}. Photogenerated holes build-up at the interface due to the presence of the interfacial oxide layer \cite{song2015role}. The oxide layer poses a tunneling barrier for charge carriers and an electric field facilitates the tunneling of charge carriers through this barrier. This electric field across the oxide layer can only partially be provided by the externally applied reverse bias. The voltage drop within a GIS- or more generally any metal-insulator-semiconductor (MIS)-Schottky diode under reverse bias consists of the voltage drop across the interfacial oxide $V_{\rm{oxide}}$ and the depletion region in the silicon substrate $V_{\rm{depletion}}$. It can be described by \cite{princsemi}

\begin{equation}
\phi_{\rm{bi}} + V_{\rm{b}} = \underbrace{\frac{q N_{\rm{d}} x_d^2}{2 \epsilon_0 \epsilon_{\rm{Si}}}}_{V_{\rm{depletion}}} + \underbrace{\frac{q N_{\rm{d}} x_{\rm{d}} t_{\rm{ox}}}{\epsilon_0 \epsilon_{\rm{Ox}}}}_{V_{\rm{Oxide}}}
\label{eq:oxdrop}
\end{equation}

with $t_{\rm{ox}}$ the interfacial oxide layer thickness. Solving eqn.\ref{eq:oxdrop} for $x_{\rm{d}}$ allows determination of $V_{\rm{oxide}}$. Due to the low doping of the silicon substrate most of the voltage drop occurs across the depletion region and $V_{\rm{oxide}}$ is less than 10mV across an oxide layer of thickness $t_{\rm{ox}}$ = 2nm at a reverse bias voltage of $V_{\rm{b}}$ = 1.5V in the dark, equating to an electric field of $\approx$ 5 $\times$ 10$^{4}$ V/cm.

Under illumination, the photogenerated holes will move towards the interfacial oxide layer where they accumulate before they can tunnel through the interfacial layer \cite{song2015role,ng1980asymmetry}. The build-up of these charges leads to an additional electric field across the interfacial oxide layer that drives the tunneling process of charge carries through the oxide layer. The tunneling current is exponentially dependent on the voltage drop across the interfacial oxide layer \cite{ng1980asymmetry} created due to the photogenerated charges. Qualitatively, lower light intensities lead to a smaller number of photogenerated carriers and reduced additional voltage drop $V_{\rm{oxide}}$, resulting in smaller photogenerated tunneling current compared to higher light intensities. Once a threshold of hole density and thus electric field is overcome, hole tunneling through the oxide layer is enhanced. Corresponding current flow leads to a reduction of accumulated holes and a balance between the accumulated holes at the interface and the tunneling current establishes. This build-up of holes at the interface explains observed reduced photocurrent responsivity $\mathcal{R}_{\rm{c}}$ (fig.\ref{fig:pdpar}b)) and increased photovoltage responsivity $\mathcal{R}_{\rm{v}}$ (fig.\ref{fig:pdpar}c)) for low light intensities. We would like to point out the influence of the light wavelength on the reduction of photocurrent responsivity $\mathcal{R}_{\rm{c}}$. Light of longer wavelength shows a more pronounced reduction in photocurrent responsivity $\mathcal{R}_{\rm{c}}$ for low light intensities (fig.\ref{fig:pdpar}b)). While this effect is not fully understood, we speculate that this reduction originates from the lower excess energy of photogenerated electron-hole pairs above the valence and conduction band edges in silicon for longer wavelength compared to shorter wavelength light. This reduced excess energy translates to an effectively increased tunneling barrier height for charge carriers tunneling through the interfacial oxide layer and reduced current flow.

\begin{figure}[htbp]
\centering{
\includegraphics[width=80mm]{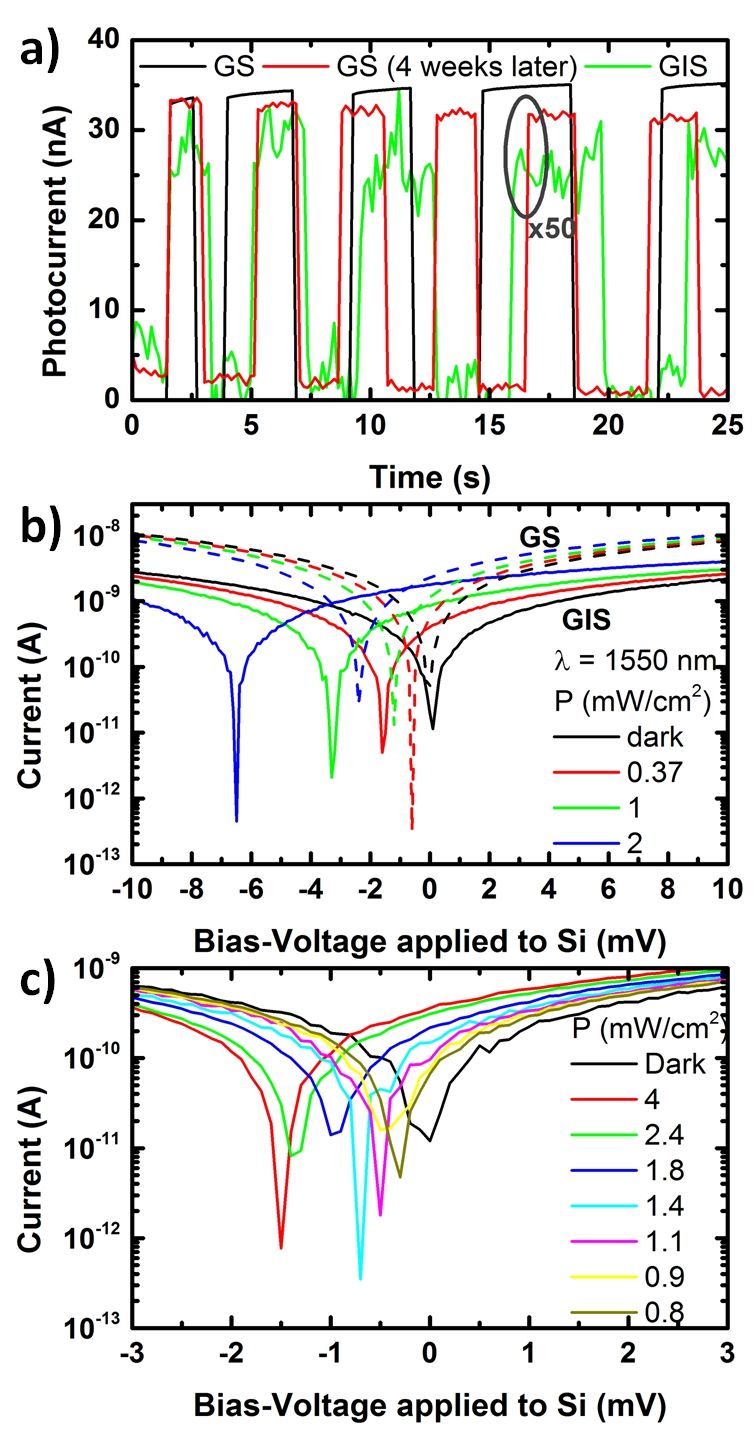}}
\caption{a) Time traces of the photocurrent of the GIS and GS devices under illumination with light of wavelength $\lambda$ = 1.55$\mu$m. b) Current-voltage dependence of the GIS and GS device under illumination with light of wavelength $\lambda$ = 1.55$\mu$m. c) Current-voltage dependence of the GIS device under broadband thermal irradiation of a hot body (T = 550K) at different radiation intensities.}
\label{fig:ir}
\end{figure}

The interfacial oxide layer further influences the light detection capabilities of graphene-silicon Schottky diodes in the infrared wavelength regime beyond 1.1$\mu$m where graphene is the main light absorbing material. Fig.\ref{fig:ir}a) shows the time dependent photocurrents under periodic illumination with light of wavelength $\lambda$ = 1.55$\mu$m of the GIS and GS device. In its pristine state, the GS device exhibits photocurrents of 35nA (R = 0.3 mA/W). The photocurrent reduces by more than 50 times after four weeks of aging and corresponding re-growth of the interfacial oxide layer in the GS device and increase in SBH, comparable to the GIS device. However, despite the reduced photocurrent response in the presence of an interfacial oxide layer, the full current-voltage curve in fig.\ref{fig:ir}b) shows that the GIS device exhibits an increased photovoltage response V$_{\rm{oc}}$ compared to the aged GS device. Further, the GIS device exhibits a photovoltage response under illumination with thermal radiation of a hot body heated to temperature T = 550K, incident on the device filtered with a $\lambda$ = 1.5$\mu$m long pass filter (\ref{fig:ir}c)). The interfacial oxide layer and corresponding increased SBH is as such not detrimental for light detection in both the visible and infrared wavelength ranges. Instead, depending on the required application and utilized read-out mechanism (current vs. voltage), the interfacial oxide layer can enhance the photodetection properties of graphene-silicon Schottky photodiodes.

\begin{figure}[htbp]
\centering{
\includegraphics[width=80mm]{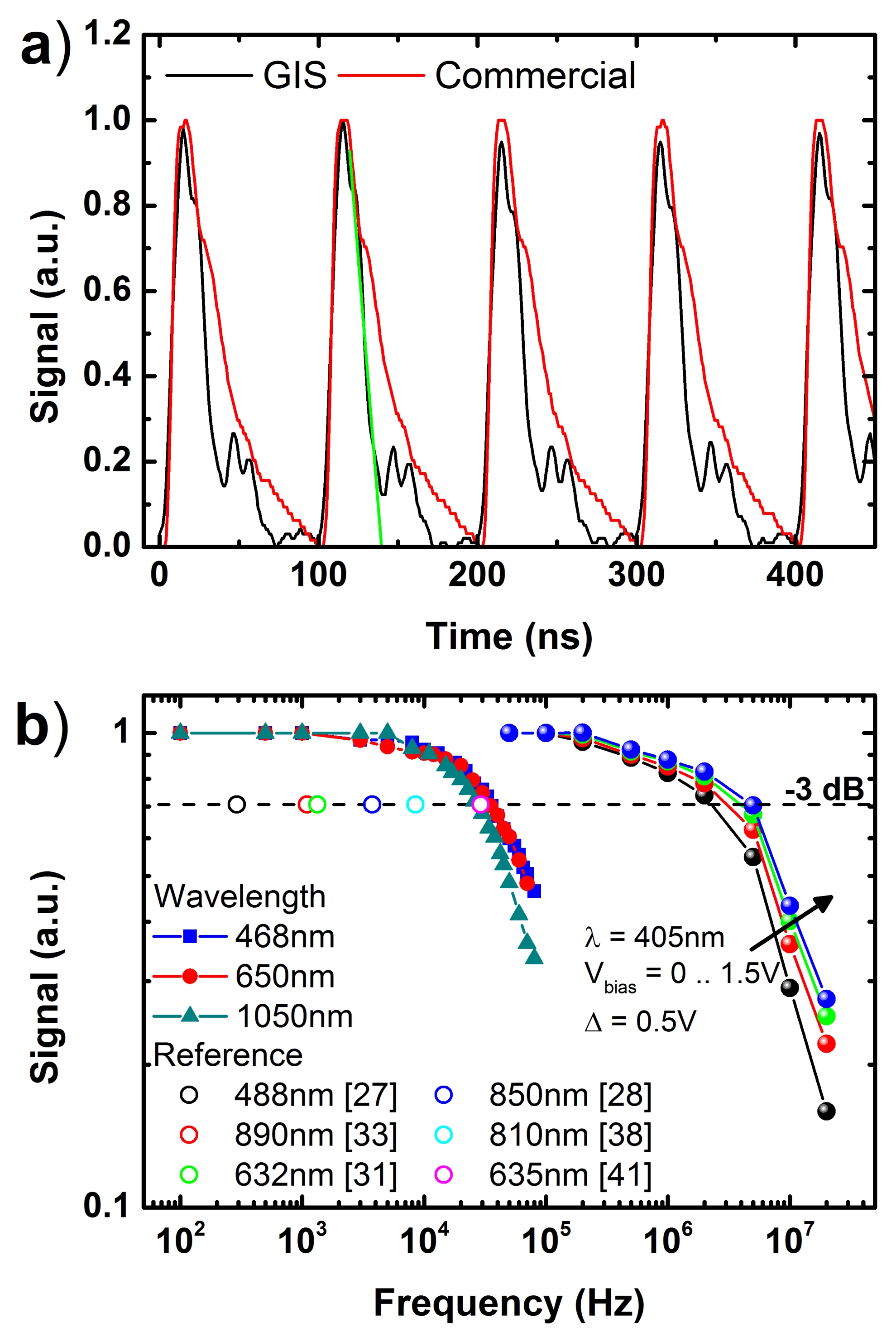}}
\caption{a) Photocurrent response of the GIS device and comparison with a commercial photodetector under illumination with laser light of 10MHz repetition rate and 80ps pulse length at a wavelength of $\lambda$ = 405nm. The decaying part of the GIS device response curve has been fitted to exclude ringing effects attributed to our band width limited setup (green line). b) Frequency dependent response of the GIS device at different illumination wavelengths and comparison with graphene-silicon Schottky photodiodes reported in the literature.}
\label{fig:speed}
\end{figure}

The high-speed performance of the GIS device has been investigated using a $\lambda$ = 405nm laser source with 80ps pulse width and variable repetition rate. A wavelength of $\lambda$ = 405nm was chosen to reduce response speed limiting diffusion currents in the device. Fig.\ref{fig:speed}a) shows that the GIS device is faster than a commercial photodetector (Thorlabs DET210/M) of comparable active device area. The recorded time dependent response of the GIS device further exhibits ringing when the signal decays which we attribute to the bandwidth limited (50MHz) amplifier in our experimental setup. The rise- and fall-times (10-90\%) of the GIS device are 12 and 20ns, respectively. Fig.\ref{fig:speed}b) shows the frequency dependent response at different illumination wavelengths. The cut-off frequency f$_{\rm{c}}$ can be determined from the drop in signal amplitude by 3dB. The GIS device exhibits a cut-off frequency f$_{\rm{c}}$ = 2..5MHz at a wavelength $\lambda$ = 405nm that can be increased with increasing reverse bias voltage due to an increase of the depletion region length and corresponding reduced capacitance of the junction. For longer wavelengths, f$_{\rm{c}}$ drops to $\approx$ 27..35kHz. Further, 
the cut-off frequency f$_c$ exhibits a wavelength dependence, it reduces with increasing wavelength from f$_{\rm{c}}$ = 35kHz at $\lambda$ = 468nm to f$_{\rm{c}}$ = 27kHz at $\lambda$ = 1050nm. This decrease in cut-off frequency f$_c$ by two orders of magnitude can be explained due to a strongly increased contribution of diffusion currents to the overall photoresponse of the device for wavelengths longer than 400-500nm, depending on the depletion region length (fig.\ref{fig:theo}b)). Especially in low-doped silicon substrates, long carrier lifetimes in the order of tens to hundreds of $\mu$s that dominate diffusion currents limit the high-speed photoresponse of devices. As reference, the cut-off frequency of various device reported in the literature have been included. The cut-off frequency f$_{\rm{c}}$ for these reference devices has been determined from the reported rise-time $t_{\rm{rise}}$ as f$_{\rm{c}}$ = 0.34/$t_{\rm{rise}}$ \cite{liuphotonic}. To the best of our knowledge, the reported rise-/fall-times and f$_{\rm{c}}$ of our devices are the fastest reported for a graphene-silicon Schottky photodiode operating in the visible wavelength range to date. The cut-off frequency of the device can potentially be further increased by reducing the area of the device to decrease its capacitance, e.g. the area of the shown device of 60mm$^2$ can be straightforwardly reduced by several orders of magnitude. Reducing the thickness of the substrate and engineering the doping profile of the silicon substrate to achieve e.g. a n$^-$-n$^+$ doping profile can reduce response-speed limiting diffusion currents in devices.  It should be noted that in the wavelength region beyond 1.1$\mu$m where silicon is not light absorbing, devices are able to operate at high speeds due to reduced diffusion currents in the silicon substrate.

In conclusion, we demonstrated the influence of the silicon substrate and its interfacial oxide layer on the properties of graphene-silicon Schottky photodiodes. Consideration of drift and diffusions currents in the substrate upon light absorption and their dependence on the depletion region length allows modeling the spectral sensitivity of the photodiode and optimizing devices by engineering doping level and profile as well as thickness of the substrate. We showed that the interfacial oxide layer leads to an increase in SBH height and reduces leakage currents under reverse voltage bias of the diode. The interfacial layer re-grows with increasing lifetime of devices, a factor that needs to be addressed through e.g. appropriate passivation in future devices that require low SBH. While the interfacial oxide layer increases the SBH and leads to a reduction in photocurrent responsivity, we demonstrated that this interfacial layer leads to an increase in photovoltage responsivity, particularly for low light intensities. Both broad spectral photodetection range from the near-UV to the short-/mid-IR wavelength regime and high-speed light detection capability with rise- and fall-times in the order of tens of ns despite large junction area demonstrate the potential of graphene-silicon Schottky photodiodes. Further optimization of the substrate, e.g. dopant profile engineering and tailoring of substrate and interfacial layer thickness will further pave the way for graphene-silicon Schottky photodiodes towards real-world applications.

\section{Acknowledgements}

H.S. acknowledges funding from the Turkish government (MEB-YLSY) and N.U. acknowledges funding from The Royal Thai Army. P.P. acknowledges funding from the Royal Society (RG140411). This work was partially funded by EPSRC contract EP/P015581/1. T.J.E would like to thank K.S. Novoselov for providing experimental equipment.

\end{document}